# Photophoresis in Single Walled Carbon Nanotubes

Gopannagari Madhusudana,[1] Vikram Bakaraju[1] and H. Chaturvedi[1,a]

[1]*Dept. of Physics, Indian Institute of Science Education & Research, Pune 411008*

**Abstract:**

We report specifically two things, one the phenomenon of optically induced motion in pristine single walled carbon nanotubes (SWNT) on the macro-scale and other the consequent separation of metallic and semiconducting enriched SWNT aggregates. Experiments provide direct evidence of both positive and negative photophoresis of SWNTs in solution *i.e.* motion away and towards light respectively. This optically induced motion was found to be dependent on frequency and intensity of light. Aggregates of pristine SWNT, moving under UV and Visible lamp were separated and characterized using absorption and Raman spectroscopy. Aggregates separated from pristine SWNT show enrichment in metallic or semiconducting SWNTs, depending on the spectral frequency of the lamp. Photophoresis in selective SWNTs show direct relation between frequency of illumination and absorption of specific nanotubes. The observed phenomenon is also verified using pre-separated, metallic and semiconducting SWNTs in solution. Metallic SWNTs show enhanced photophoresis with an UV lamp, whereas pre-separated semiconductor SWNT exhibit motion with a mercury lamp with broadband visible frequency. However, under IR lamp, both metallic and semiconducting enriched SWNT show continuous motion, as seen for un-separated pristine SWNTs, where no specificity was observed and all particles were seen moving.

[a]Electronic mail:*hchaturv@iiserpune.ac.in*

Directed motion of particles provides an important paradigm for specific functionalization and bottom up self-assembly of nanoparticles and biomolecules into functional devices. The motion of particles has been widely reported using electric, magnetic and thermal gradients.[1-3] Efficient processes of separation, selective binding in particles, advances in biological assays, and microfluidic devices are envisioned using directed motion of micro/nano particles or biomolecules.[4, 5] Photophoresis, which is optically-induced motion in particles has been studied from the observations of Ehrenfratin in 1916[6] to the motion of micron sized aerosol, carbon or smoke particles in stratospheres under various external conditions.[7, 8] The dependence of photophoresis on various parameters such as size, shape, conductivity, and refractive index, specifically in the case of metal nanoparticles has also been analytically reported.[9-11] The first report of photophoresis in an aqueous solution was by Barkas[12] in 1926, but since then, there has been limited experimental observations with respect to nanoparticles in soluion.[13, 14] Motion in single walled carbon nanotubes (SWNT) due to electro-magnetic fields, and high-energy optical traps has been reported.[15, 16] Here, we report large macro-scale motion in aggregates of pristine SWNTs by simple lamps. Aggregates of SWNTs show both positive and negative photophoresis *i.e.* motion away or towards the light source. Our experiments distinctly demonstrate controlled directed motion of the aggregates of pure pristine SWNTs. Under uniform UV or visible illumination, SWNTs show distinct separation of the aggregates, and motion is observed only in part of the aggregates while rest remains at the bottom of the vial. Separated particles show enrichment in metallic SWNTs, whereas enhanced motion and enrichment is observed for semiconducting SWNTs under visible illumination. Photophoresis in selective SWNT show direct relation between the spectral frequency of illumination and dielectric property of the SWNTs used, whether metallic or semiconducting. Separation of semiconducting and metallic

SWNT is an area of active research.[17, 18] We believe optically directed motion in absorbing SWNT may lead to a non-surfactant and non-destructive process of separation of SWNT of specific diameters. Considering both photophoresis and nanoparticles has wide implications across diverse fields such as atmospheric sciences, colloidal and nanosciences[19, 20]; we believe optically-induced directed motion in SWNT as reported here, may be important for diverse applications such as micro-fluidics, optical sensors, actuators and future technologies requiring stability and separation of nanoparticles.

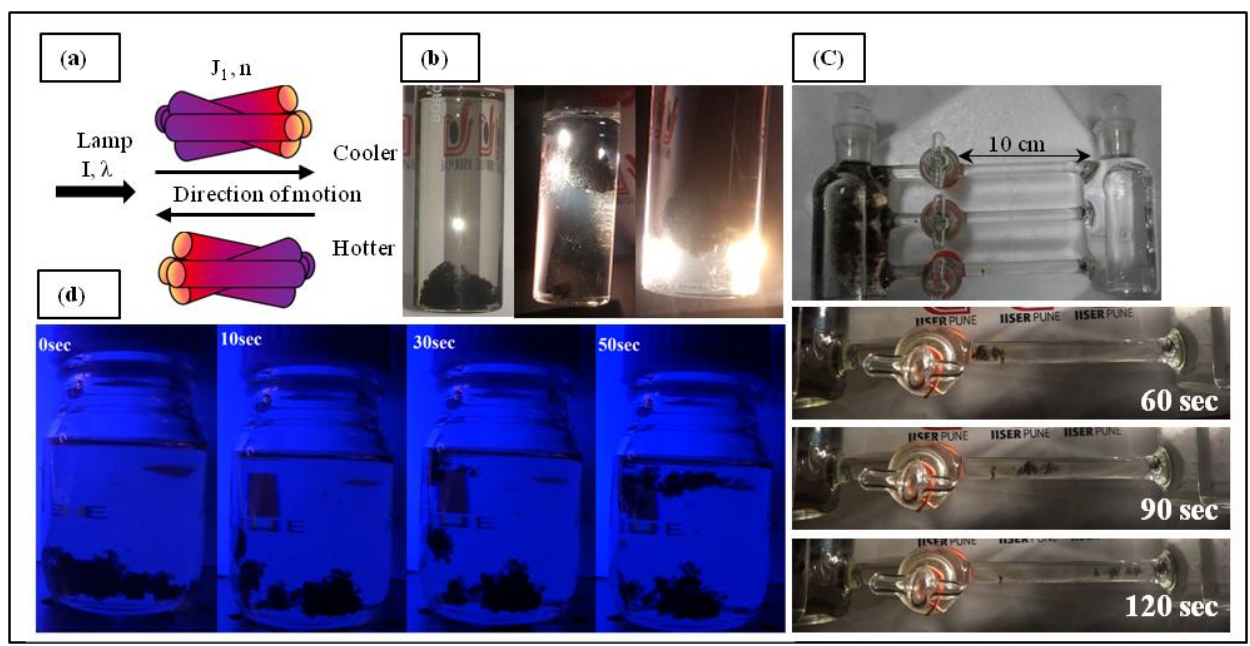

FIG. 1. (a) Conceptual model showing temperature difference (ΔT) in aggregate of SWNTs in light with intensity(I) and wavelength(λ), inducing motion. (b) Controlled motion of aggregated pristine SWNT's with directed beam of light. (c) show timeline of negative photophoresis in aggregated SWNT's. (d) Timeline of optical induced motion and distinct separation of specific nanotubes under UV light.

Light induced motion in particles occurs due to the temperature difference (ΔT) on the surface of particle. If the surface exposed to the light gets hotter than the surface facing away from the light, then the particle moves away from light as positive photophoresis. However, in certain cases depending on particle size and refractive index, the surface facing away from light may be hotter than the surface towards light. In such cases, negative photophoresis is observed and particles are directed towards light.[21] Photophoresis in individual cylindrical particle is caused by the localized, non-uniform heat distribution on the particle.[22] A particle absorbing light at higher temperature than the surrounding gas will experience photophoretic force depending on the accommodation coefficient Δα.[23] Photophoretic velocity for a cylindrical particle subjected to an electric field, $E_0$, is given as $U = -\frac{C_s J_1}{2(1+2C_m^*)(1+\kappa^*+k^*C_t^*)}\frac{\eta I}{\kappa \rho T_\infty}$. Where, photophoretic asymmetry factor $J_1 = \frac{4\pi \upsilon \kappa a}{\lambda}\int_0^1 \int_0^{2\pi} B(\zeta,\phi)\zeta^2 \cos\phi \, d\phi \, d\zeta$ depends on the complex refractive index (n = $\upsilon + i\kappa$) and the normalized size ($2\pi a/\lambda$) of the particle,[21, 24] λ being wavelength of incident radiation and $B(\zeta,\phi)$ is the dimensionless electric field distribution. Hence, direction of motion significantly depends on dielectric response of the particle to the optical excitation. Depending on the asymmetry factor ($J_1$), photophoretic velocity may be positive or negative for an individual cylindrical particle absorbing light. Numerical results for optically-induced forces in aggregates of metal nanoparticles also show dependence on plasmonic resonance frequencies.[10] Hence, photophoresis in nanoparticles such as either metallic or semiconducting SWNTs is expected to depend on the characteristic plasmonic or absorption frequency respectively, *i.e.* on complex refractive index or dielectric property of the particle. Pristine pure SWNTs, as produced are mixture of nanotubes with both metallic and semiconducting properties. Aggregates of pure SWNT have nanotubes of various diameters and dielectric properties, signifying different absorption coefficients and refractive indices. Hence, when these aggregates of pristine SWNT are subjected to resonant optical frequency, motion in absorbing particle is observed due to significant, induced photophoretic forces, as shown in Fig.1.

Pristine SWNTs (purity 99.8 %) were used, as purchased from Nanointegris, without any further modification or purification. In this report we have only considered pure SWNTs in solution of DMF; however we also observe similar optically induced motion of SWNT in other organic solvents like acetone and chloroform.

Fig 1 b. show controlled motion of the SWNT aggregates as directed by the optical beam. As shown, aggregated floc can be moved either to the top or to the bottom of the vial, controlled by the narrow beam of light. Photophoresis in these aggregates of SWNT show a complex dependence on the intensity of the beam. Samples were illuminated using optical fiber coupled through the mercury lamp. At lower intensities (10-20 $J/m^2$) particles moves slowly towards the light source and remains in the inclined beam flux, however for higher intensities (140-220 $J/m^2$) same aggregates show motion away from the central flux of the beam. Particles show rotating motions there by dissipating energy, while the aggregated floc as a whole remains in the directed narrow beam. Although experimental observations confirm controlled, directed photophoretic motion, care was taken so as to minimize the theromophoretic effects in solution if any. The solvent DMF also do not absorb in the UV or visible frequencies used for illumination, thus negating any doubts of the thermal convection due to absorption of the solvent. Experiments in the specially designed glass contraption in Fig 1 (c) clearly shows negative photophoresis, a part of the aggregated floc can be seen moving over 10 cm of the channel connecting two flasks, in the direction of light in just 2 minutes of illumination. The glass contraption clearly demonstrates potential application of the phenomenon for optically directed, controlled transportation of nanoparticles functionalized by various molecules, as required by biological or microfluidic applications and separation technologies.

To further probe the dependence of the phenomena on the frequency of illumination, aggregated SWNTs were uniformly illuminated from top, using commercially available UV (125W, 352 nm) and broadband mercury spectral lamp (80W). Under uniform optical illumination, part of the aggregated pristine SWNT in DMF solution moves to the top of the vial, while rest remains at the bottom. As shown in the Fig.1 (d), within 50 seconds particles moves to the top of the vial under UV lamp and clearly separates out from the rest. Part of the aggregate of SWNT under visible lamp also shows similar motion and distinctly separates out as shown in Fig 2 (a). However, aggregates under IR lamp show neither any distinct separation nor directed motion. Unlike the distinct separation under UV or visible illumination, showing motion only in selective part of the aggregated flocs; rapid, continuous motion in complete aggregated floc under NIR lamp is observed across the vial within 10 seconds (check video clip as supplementary Information).

SWNTs moving to the top under lamps were separated from the ones at the bottom of vial and characterized using UV-Vis-NIR absorption spectrometer and Lab Ram Raman spectrometer with 632 nm laser line. In case of SWNTs the optical absorption significantly depends on the diameter of the nanotube.[25] Absorption of pristine one-dimensional carbon nanotubes differ significantly from bulk material with its characteristic van hove singularities. The transitions are relatively sharp and characteristic to specific diameter and band gap energies of semiconducting nanotubes. However, significant absorption by pristine SWNTs in UV is essentially due to the plasmonic resonance of metallic nanotubes. Hence absorption spectra of the pristine SWNT show characteristic features of both metallic $M_{11}$ (350-600 nm) and semiconducting inter-band transitions $S_{11}$ (900-1400 nm) and $S_{22}$ (550-900 nm) as shown in Fig.2(c, d). The SWNTs collected from the top of the vial under UV lamp, shows enhanced absorption in the plasmonic metallic $M_{11}$ band; as also seen for SWNTs collected from bottom of the vial, under visible frequency lamp. Similarly, nanotubes separated from the top of vial under mercury lamp and bottom of vial under UV lamp, show similar and significant increase in absorption at visible NIR frequency. Since NIR frequency is directly related to the band gap absorption by semiconducting tubes, increase in NIR absorption in separated aggregates indicate enrichment of semiconducting SWNTs.

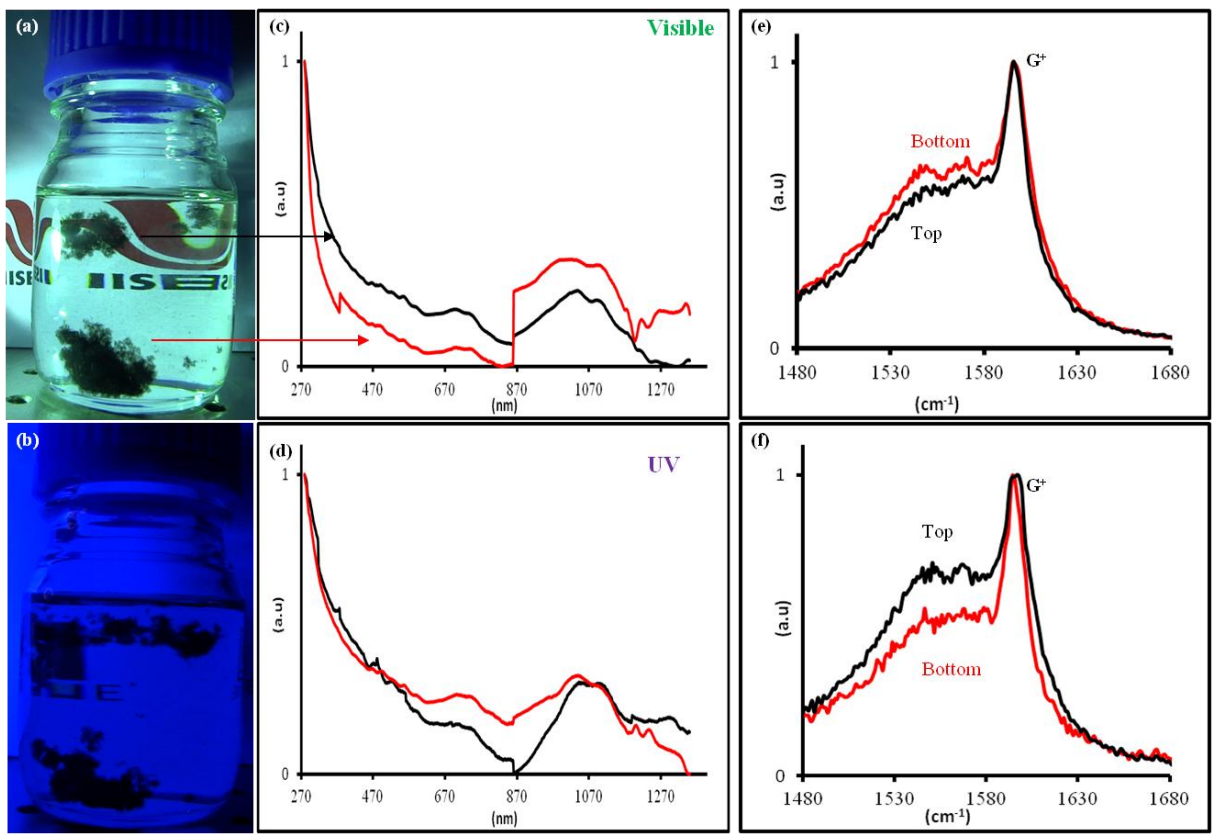

FIG. 2. (a, b) Optically induced motion in selective pristine SWNTs and consequent separation under visible frequency mercury and UV lamps. (c, d) Normalized absorption spectra of pristine SWNT aggregates separated from bottom (red) and top (black) of vials under mercury and UV lamps. (e, f) The G-band of raman spectra aggregates pristine SWNTs shows motion and enrichment of metallic and semiconducting nanotubes at the top of vial under UV and mercury lamps.

Further characterization by Raman spectroscopy confirms this enrichment in separated SWNTs, exhibiting optically induced motion and switching depending on the frequency of illumination. Analysis of the Raman spectra of SWNT show two main components of the $G$ band, as shown in Fig.2(c), one at 1592 cm$^{-1}$ ($G^+$) and the other broadly peaked at 1568 cm$^{-1}$ ($G^-$).[26] The $G^+$ feature is associated with vibrations of carbon atom along the nanotube axis (LO phonon mode) and the $G^-$ feature, is associated with vibrations along the circumferential direction of the SWNT (TO phonon). The $G^-$ is highly sensitive to the metallic to semiconducting ratio in pristine SWNT aggregates. Metallic enriched SWNT show Breit–Wigner–Fano line shape; whereas lorentzian line-shape is observed for semiconducting enriched SWNTs. An increase in $G^-/G^+$ ratio, and corresponding broadening suggests increase in metallic SWNTs,[27] as shown in Fig.2 (e, f). Broadening and changes in relative intensity of the $G^+$, $G^-$ band for pristine SWNT collected from the top and bottom of the vial under UV and visible lamps; verifies enrichment of specific metallic and semiconducting SWNT, as was indicated by similar changes in either the plasmonic $M_{11}$ or semiconducting $S_{11}$ band of the absorption spectra. Both samples, ones separated from the top of the vial under UV illumination and the other from the bottom of vial under visible lamp; show similar increase in $G^-/G^+$ band ratio. In contrast, the ones separated from the top under visible illumination show similar enrichment as ones from the bottom of the vial under UV lamp.

Additionally, the radial breathing mode (RBM) of the separated nanotubes shows enrichment of specific diameter tubes due to motion under UV and mercury irradiation. Fig.3(a, b) show the RBM of pristine SWNTs separated under UV and visible illumination. The frequency of the RBM is inversely proportional to the diameter of individual SWNT.[28]

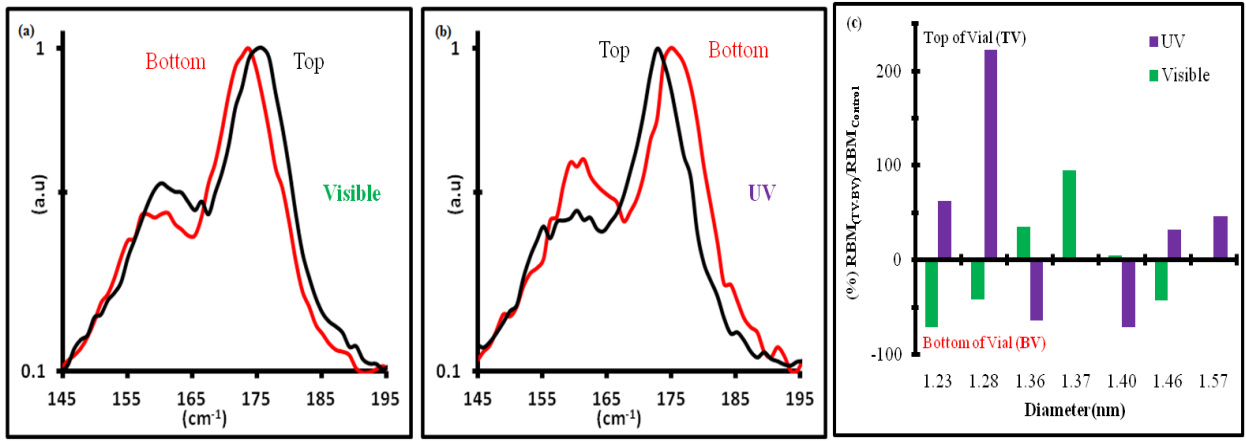

FIG. 3. (a, b) Radial Breathing Mode (RBM) showing enrichment of specific diameter of SWNTs under mercury and UV lamps. Aggregates of SWNTs separated from the top (black) under mercury lamp and bottom (red) under UV (Black) lampshow similar enrichment. (c) histogram as plotted from the RBM data showing enrichment in specific diameter of the SWNTs at the top or bottom of the vial due to motion under visible mercury lamp (green) and UV lamp (violet).

The RBM of each sample was fit using multiple lorentzian functions to identify relative enrichment in individual diameter tubes in the Raman spectra of pristine SWNT. The intensity of the RBM in each separated SWNT sample was normalized with respect to similar diameter tubes fit in the un-separated pristine SWNT sample that was used as a control. The diameter of the SWNT was calculated using $\omega_{RBM} = (\alpha_{RBM}/d) + \alpha_{bundle}$.[29] Where d is the diameter of nanotube and $\alpha_{bundle}$, $\alpha_{RBM}$ are constants associated with bundling effect and scaling factor in RBM spectra of nanotubes respectively. Individual RBM of each sample was fitted and normalized using multiple lorentzian peaks, representing different diameter of SWNTs. Corresponding histogram Fig.3(c)is plotted using RBM spectra of separated SWNTs. It shows relative enrichment of specific diameter SWNT and the direction of motion with respect to un-separated pristine-SWNTs. The discernible changes in the RBM and the G band of the vibrational spectra provide further evidence enrichment of specific nanotubes; due to optically induced motion. The motion of SWNT aggregates under UV illumination conclusively show enrichment in metallic SWNT with specific diameters. Similarly, under visible mercury lamp, SWNT aggregates moving to the top of vial are found enriched in specific diameter of semiconducting SWNTs.

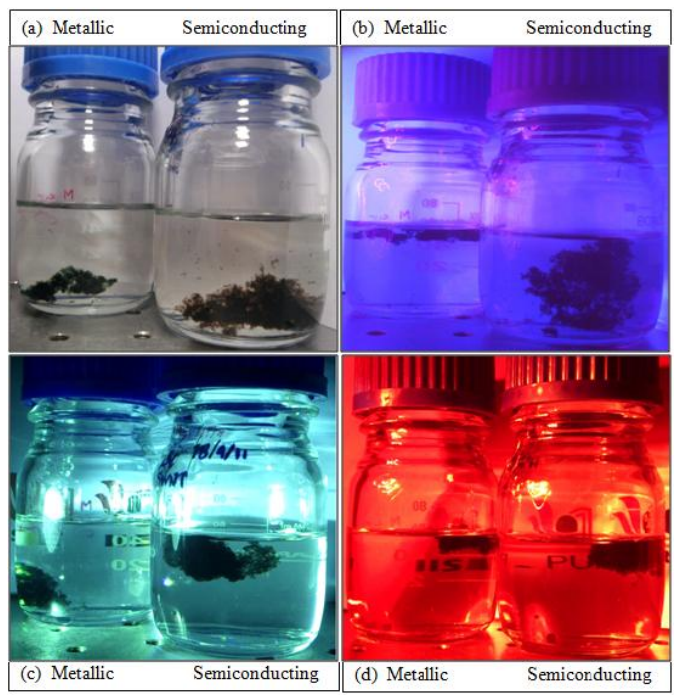

FIG. 4. Photophoresis in aggregates of pre separated (95% pure) metallic (left) and semiconducting (right) SWNTs. (a) Aggregates at bottom of vials in dark. (b) show optically induced motion in metallic SWNTs under UV lamp. Whereas (c), shows enhanced motion of semiconducting SWNTs under visible frequency mercury lamp. However as shown in (d) both metallic and semiconducting SWNTs show enhanced motion under IR lamp.

Photophoretic separation and consequent enrichment of metallic and semiconducting SWNTs from the pristine SWNTs was further verified using pre-separated semiconductor, metallic enriched solutions of SWNTs. Pre-separated semiconductor and metallic SWNT were purchased from Nanointegris (with 95% Purity) and used without any further modification. As shown in Fig.4(a-d), samples of semiconductor and metallic enriched aggregates show complimentary motion under UV and broadband mercury lamp. However, in case of illumination by IR lamp, there is no specificity in direction of motion or in diameter of nanotubes. Solutions of both metallic and semiconducting SWNT show continuous motion under IR lamp, as observed in case of pristine SWNTs. Photophoresis in pre-separated metallic or semiconducting SWNTs, along with comparative enrichment shown in the absorption and Raman spectra of separated pristine SWNT indicates photophoresis in SWNT aggregates depends on the respective absorption either by metallic or semiconducting SWNTs under UV and mercury lamps.

Large scale, optically controlled motion and consequent separation of SWNTs opens up an interesting areas, on one hand it potentially provides solutions for issues of active concern such as sepration and controlled transportation in SWNTs on other hand it challenges conventional fundamental understanding seeking and active research. Motion in selective SWNTS based on frequency of simple lamp, proposes to be an efficient way for both large scale and specific separation of SWNTs. Since, motion in SWNT depends on dielectric response we believe the phenomenon should be useful for directed motion and efficient separation in other nanoparticles too. Photophoresis in SWNT aggregates presents an exciting avenue for optically controlled motion and induced separation in other nanoparticles such as metallic nanoparticles and quantum dots based on either their size or electro-optical properties. Along with its applications in various advanced technologies, SWNT also provide an opportunity to understand novel effects in colloidal solutions of one-dimensional nanoparticles. Clearly both theoretically and experimentally much research is needed, to fully understand the phenomenon of photophoresis in the aggregates of one dimensional particles like SWNTs, having different diameters, with distinct optical and dielectric properties. However, in this study, we experimentally demonstrate and attempt to provide possible theoretical reasons for optically induced motion especially negative photophoresis and enrichment in separated SWNTs. In conclusion, we propose this optically controlled motion and subsequent separation of SWNT in solution may play an important role for directed self-assembly and technological advances in diverse fields such as molecular devices, micro-fluidics, biotechnology and industrial applications requiring separation of nanoparticles.

**Acknowledgments**

Authors are deeply indebted to Dr. Jordan C. Poler and Andrea Giordano for invaluable scientific input, Ramanujan fellowship (SR/S2/RJN-28/2009) and funding agencies DST (DST/TSG/PT/2012/66), Nanomission (SR/NM/NS-15/2012) for generous grants.